\documentclass[prb,preprint,superscriptaddress,showpacs,amsmath,amssymb]{revtex4}

\usepackage{graphicx}
\usepackage{dcolumn}
\usepackage{bm}

\begin{document}

\title{Physical mechanisms of nonlinear conductivity: a model analysis}


\date{\today}

\author{Andreas Heuer}
\affiliation{\frenchspacing Westf\"alische Wilhelms Universit\"at M\"unster, Institut f\"ur physikalische Chemie, Corrensstr.\ 30, 48149 M\"unster, Germany}
\affiliation{\frenchspacing Center of Nonlinear Science CeNoS, Westf\"alische Wilhelms Universit\"at M\"unster, Germany}
\author{Lars L\"uhning}
\affiliation{\frenchspacing Westf\"alische Wilhelms Universit\"at
M\"unster, Institut f\"ur physikalische Chemie, Corrensstr.\ 30,
48149 M\"unster, Germany} \affiliation{\frenchspacing Center of
Nonlinear Science CeNoS, Westf\"alische Wilhelms Universit\"at
M\"unster, Germany}


\begin{abstract}
Nonlinear effects are omnipresent in thin films of ion conducting
materials  showing up as a significant increase of the
conductivity. For a disordered hopping model  general  physical
mechanisms are identified giving rise to the occurrence of
positive or negative nonlinear effects, respectively. Analytical
results are obtained in the limit of high but finite dimensions.
They are compared with the numerical results for 3D up to 6D systems.
A very good agreement can be found, in particular for higher dimensions.
The results can also be used to rationalize previous numerical
simulations. The implications for the interpretation of nonlinear
conductivity experiments on inorganic ion conductors are
discussed.
\end{abstract}






\maketitle

\section{Introduction}

The preparation of thin films of solid ionic conductors  helps to
improve their application in micro-batteries, fuel cells, and
electrochromic devices, see, e.g. \cite{Croce, Noda}. The
resulting large electric fields give rise to nonlinear
contributions for the conductivity which in practice turn out to
be positive \cite{Isard96,Murugavel05,Staesche} thus improving the
applicability of, e.g.,  micro-batteries.

From a theoretical perspective the ion dynamics in disordered
inorganic ion conductors is a complex multi-particle problem
\cite{ingram:1987,dyre:2009} with strong interactions among the
mobile cations \cite{maass:1999,Funkerecent}. Interestingly,
several features of the conductivity have been reproduced by
analyzing a single-particle hopping motion in a discrete
disordered energy landscape
\cite{hunt:1991,baranovskii:1999,dyre:2000,dyrenew}. Recently this
approach has found a numerical justification since to a good
approximation the ion dynamics can be mapped on a single-particle
vacancy dynamics between distinct sites \cite{lammert:2010}. Also via molecular dynamics simulations
key properties of the nonlinear conductivity can be gained for realistic
microscopic systems such as alkali silicate systems\cite{kunow:2006}.

Model energy landscapes are frequently used to study transport
phenomena such as anomalous diffusion, see, e.g.,
\cite{Ambegaokar69,Derrida83,Bouchaud90,Khoury}. Furthermore, for
1D lattice as well as continuous models it is possible to derive
exact expressions for the field-dependent current
\cite{Kehr97,Heuer:05, Eimax}. Of major current interest is also
the study of driven open systems; see, e.g., Ref.\cite{Maass12}.
Interestingly, for periodic boundary conditions the conductivity
displays non-analytic behavior at zero field in the thermodynamic
limit \cite{Roling:08,Eimax}. The numerical results indicate that
the non-analytic behavior disappears in higher dimensions so that
the 1D solution is of no relevance for the experimentally relevant
2D or 3D case \cite{Heuer:05}. Here to the best of our knowledge
no exact solutions exist which describe the field-dependence of
the dynamics in a disordered energy landscape. Formal solutions
have been formulated for general dimensions, which so far could
only be evaluated for the 1D case \cite{Igor}. For electron
hopping transport in disordered semiconductors the population of
states can be characterized by a field-dependent effective
temperature \cite{new1,new2}. Due to the different physical
background (different hopping distances, multi-particle effects)
this solution cannot be directly translated into the present
problem of interest. A general introduction in the hopping transport
in disordered systems can be found in \cite{boettger}.

Without energetic disorder the field-dependence of the current in nearest-neighbor
 hopping models with point-like sites and the electric field along
 a coordinate axis
reads $j(w) \propto \sinh(w)$ where $w= q \beta a E/2$ is the
normalized electric field (q: charge, $\beta$: inverse
temperature, $a$: hopping distance, $E$: electric field)
\cite{Maier}. More generally one expects for isotropic systems
\begin{equation}
j(w) = \sigma_1 w + \sigma_3 w^3 + ... \quad .
\end{equation}
An appropriate dimensionless measure for the nonlinearity is given
by $\sigma_r = 6 \sigma_3/\sigma_1$. For the ordered hopping
model, introduced above, $\sigma_r$ is unity if all site-energies
are identical, i.e. for $j(w) \propto \sinh(w)$. Note that
$\sigma_r$ can be directly obtained from the experiments and
reaches values of the order of 100 \cite{Murugavel05,Staeschezpc}.
Note that nearest-neighbor hopping dynamics and the
presence of discrete sites is by far the dominant transport mechanism
in alkali silicate systems in the linear and nonlinear regime \cite{lammert:2003,kunow:2006}.

For single-particle hopping models on a lattice it is has been
shown  numerically that one
 obtains $\sigma_r < 0$ for a broad box-type distribution of
energies whereas a broad Gaussian energy distribution gives rise to $\sigma_r \gg 1$, in agreement with the experimental results
  \cite{ZPC}. Furthermore, for the Gaussian case
it has
been observed that the specific value of $\sigma_r$ is strongly influenced by the
presence of some few low-energy sites, representing the low-energy wing of
the Gaussian. However, beyond these merely numerical results
no physical understanding about the underlying
mechanisms for the origin of the nonlinear effects has been gained.

In this work we discuss a disordered hopping model in arbitrary
dimension with a bimodal energy distribution. This model is simple
enough that strict analytical results for the nonlinearity can be
obtained. However, at the same time it is complex enough that one
can clearly extract the non-trivial underlying mechanisms which are
responsible for the two different limits of either positive or
negative nonlinear behavior. In particular, the numerical results
for the box-type and the Gaussian distribution, as reported above,
can be qualitatively understood based on the  results of this work.
Furthermore, this allows one to suggest the implications of the
large positive nonlinear effect, seen experimentally. Since in the
experiments only the third-order term of the current can be
measured with sufficient accuracy we finally express our results
in this experimentally relevant limit.

More specifically, we will proceed in two steps. (1) Exact
numerical calculations are performed for dimensions between 3 and
6. (2) An analytical treatment is presented which contains the
corrections to the mean-field case (infinite dimension) as an
expansion in the inverse dimension. It thus becomes a good
description in the limit of high dimensions. From comparison of
the results of (1) and (2) it will turn out that the analytical
results already reflect quite well the numerical results for the
experimentally relevant 3D case and constitute an excellent
reproduction of the results in higher dimensions.

The outline of this manuscript is as follows. In Sect. II details
of the hopping model and the procedure for the analytical
calculations  are presented. Sect. III presents the numerical
results for the linear and the non-linear conductivity in 3D. Next, Sect. IV
contains the analytical results which are then compared with the numerical
results in Sect. V.  Finally, in Sect. VI we discuss the results.

\section{Model and Simulations}

We consider a $d$-dimensional hypercubic lattice where all sites
possess an energy, drawn from a bimodal energy distribution.
Periodic boundary conditions are employed. The two energy values
are denoted $\pm \Delta U/2$. Their distribution on the lattice is
spatially uncorrelated. The fraction of low-energy sites with
energy $- \Delta U/2$ is denoted $\alpha$. This parameter will
turn out to be the key parameter for the present analysis.
Correspondingly, the fraction of high-energy sites is $1-\alpha$.
In total, one has $N^d$ sites. The external field is applied along
one of the lattice directions. We introduce the variable $\Theta$
via $d=\Theta/2+1$, denoting the number of directions which are
orthogonal to the electric field (e.g. $\Theta(d=3)=4$). It will
serve as an appropriate expansion parameter. As a measure for the
disorder we introduce the variable $Z= \exp(\beta \Delta U/2)$.
For any adjacent pair of sites $j$ and $k$ we define the hopping
rate $\gamma_{jk}$ from site $j$ to site $k$. Specifically, we
choose $\gamma_{jk} = \gamma_0$ for transitions between equal
energies, $\gamma_{jk} = \gamma_0 Z$ for the down-rate (also
denoted $\gamma_{+-}$) and $\gamma_{jk} = \gamma_0/Z$ for the
up-rate (also denoted $\gamma_{-+}$). This choice implies that the
barrier between sites with equal energy is larger than the
downward barrier from a high-energy to a low-energy site. This is
 a reasonable assumption for complex energy
landscapes \cite{HeuerSilbey:1993a,HeuerArtikel}. For reasons of
simplicity we choose $\gamma_0=1$ in our analysis.

For hopping processes parallel or anti-parallel to the electric
field the jump rates are supplemented by a factor $\exp(w)$ or
$\exp(-w)$, respectively. In what follows we use the abbreviation
$ y = e^w $. For finite $w$ the stationary population probability
$q_j(w)$ in general differs from the Boltzmann distribution
$p_{j,eq}$. From knowledge of the population probabilities the
total current can be calculated as $ j(w) = \sum_{j} q_j(w) (y
\gamma_{jk} - (1/y) \gamma_{ji})$ where $k$ denotes the neighbor
site of site $j$ parallel to the direction of the field and $i$
anti-parallel to this direction.

The numerical analysis boils down to the determination of the
stationary solution of the corresponding rate equations, i.e. to
solving large systems of linear equations. In practice, we have
solved the rate equations by the Runge-Kutta algorithm together
with an adaptive step size and determined the stationary long-time
limit. The field has been applied along the direction of one axis.
For each field strength we have also inverted the electric field
so that the resulting current is an uneven function of $w$. We
have chosen system sizes $(40)^3$, $(16)^4$, $(9)^5$ and $(6)^6$,
all corresponding to a similar number of total sites. We have
checked that upon doubling the system size the values of
$\sigma_r$ change less than 3\% for $d=3$ and less than 1\% for
$d=6$ in the relevant limits of small and large $\alpha$ (see
below). For the disorder we always chose $\Delta U/k_B T= 8$.
From the resulting population probabilities
the current can be calculated. This procedure is performed for $w
= 0.1, 0.1 \cdot \sqrt{2}$, and $0.1 \cdot \sqrt{3}$. Then we
fitted $j(w)$ to a fifth-order polynomial in $w$ and extracted
$\sigma_1$, $\sigma_3$, and, correspondingly, $\sigma_r$. We
checked that a reduction of $w$ by a factor of two did not change the
results. Thus, the expansion up to the fifth order is sufficient.
The final results were averaged over 10 different random
realisations of the disorder.

\section{Numerical Results}

\begin{figure}[tb]
\centering\includegraphics[width=0.45\columnwidth]{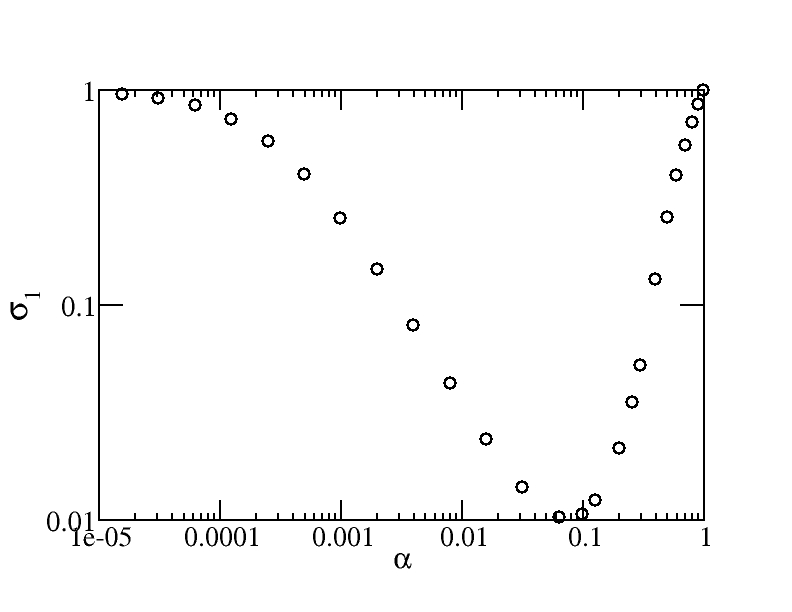}
\includegraphics[width=0.45\columnwidth]{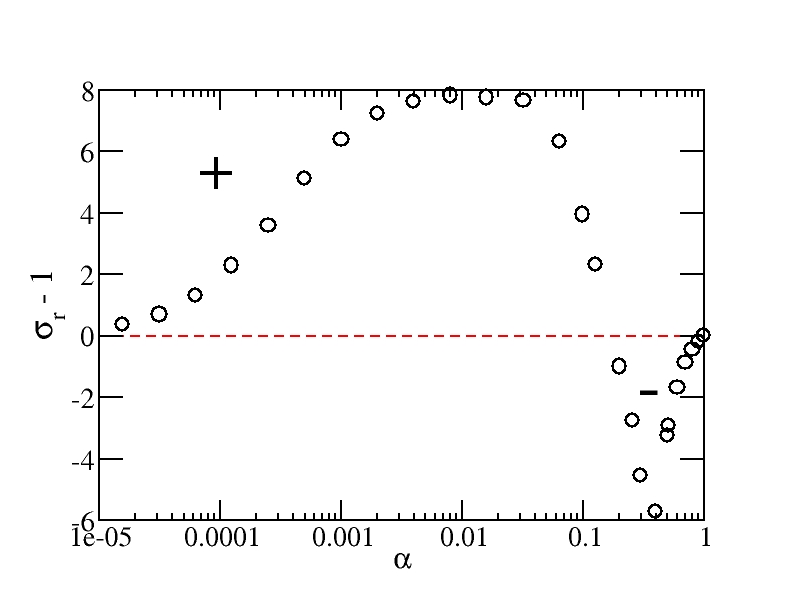}
\caption{$\sigma_1$ and $\sigma_r-1$ as a function of $\alpha$ as
obtained from the numerical simulations in 3D. $\sigma_r - 1 = 0$ is highlighted by the broken line.} \label{fig1}
\end{figure}

We start by showing $\sigma_1(\alpha)$ and $\sigma_r(\alpha)$ for
the 3D case; see Fig.\ref{fig1}. In the limits $\alpha \rightarrow 0$ and $\alpha
\rightarrow 1$ the linear conductivity is identical because the
particle is hopping in an energy landscape without any disorder.
With the chosen normalization this conductivity equals unity. The
conductivity is reduced by two orders of magnitude for
intermediate $\alpha$ with a minimum around $\alpha \approx 0.1$.
Qualitatively, the decrease of $\sigma_1$ for small $\alpha$ is
related to the fact that the number of deep-lying traps is
increasing which slows down the particle. In contrast, when
approaching $\alpha = 1$ from below, the residual high-energy
states work as barriers, slowing down the dynamics. Since the
fraction of high-energy states decreases for increasing $\alpha$
the conductivity increases correspondingly. The minimum reflects a
crossover between the trap-regime and the
barrier-regime.

Also the nonlinear conductivity as expressed via $\sigma_r$, i.e.
the ratio of the nonlinear and the linear conductivity, displays
two distinct regimes. Since for $\alpha = 0$ and $\alpha = 1$ one
has $\sigma_r=1$ by construction, we discuss the behavior of
$\sigma_r - 1$ in the following. Interestingly in the
trap-dominated regime one has $\sigma_r - 1 > 0$, i.e. an increase
of the nonlinearity as compared to the homogeneous case. In
contrast, in the barrier-dominated regime we find $\sigma_r - 1 <
0$ so that the nonlinearity is smaller. A physical understanding
of these observations will be developed in the remaining part of
this work. The simulations for higher dimensions display the same qualitative features.

\section{Analytical results}
In our analytical calculations we always consider the case of high
but finite dimensions, expressed via $\Theta \gg 1$.
\subsection{Rate equations for a single low-energy site}
\begin{figure}[tb]
\centering\includegraphics[width=0.7\columnwidth]{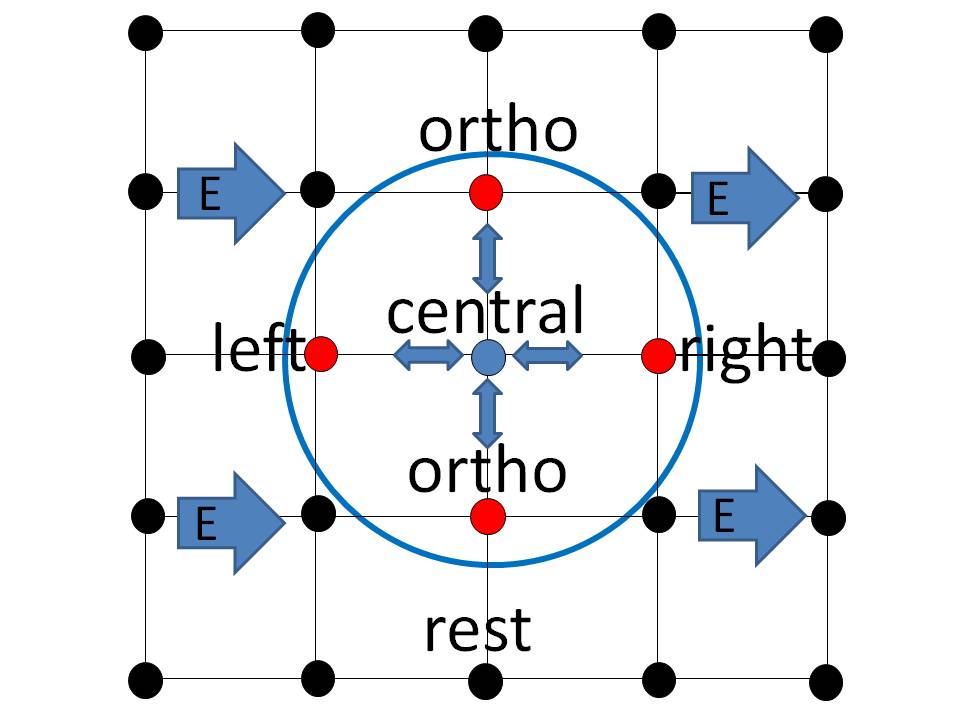}
\caption{Sketch of the notation used in the main text. If the
central site is a rare low-energy site $(\alpha \ll 1/\Theta$)
local thermodynamic equilibrium is generated, motivating the
notion of a trap-region with a border (circle).} \label{fig3}
\end{figure}

 We start with
the case of a single low-energy state. This low-energy state as well as its
direct neighbors are sketched in Fig.\ref{fig3}.  This region will
be denoted {\it trap}-region.

For setting up a set of rate equations we assume that all sites adjacent
to the trap-region possess the same non-equilibrium population $q_{rest}$ which,
of course, depends on the applied field. Later one we argue that this
assumption is not required and that it is sufficient in the limit
of high dimensions to identify $q_{rest}$ with the average population.
Then the rate equations read
\begin{eqnarray}
(d/dt) q_{central} &= & -(1/Z)(y + 1/y + \Theta) q_{central} + Z (y q_{left} + (1/y) q_{right} + \Theta q_{ortho}),
\label{rate1} \\
(d/dt) q_{ortho} &=& -(y + 1/y + \Theta - 1 + Z) q_{ortho} + (y + 1/y + \Theta - 1)q_{rest} + (1/Z)q_{central},
\label{rate2} \\
(d/dt) q_{left} &=& -(Zy + 1/y + \Theta ) q_{left} + (y +\Theta )q_{rest} + (1/(yZ))q_{central},
\label{rate3}
\\
(d/dt) q_{right} &=& -(Z/y + y + \Theta ) q_{right} + (1/y +\Theta
)q_{rest} + (y/Z)q_{central}. \label{rate4}
\end{eqnarray}
In the stationary case the left side is zero so that one
effectively obtains a system of four linear equations. In a first
step one can obtain the general equation for $q_{central}$ and
then after insertion solve for the other three populations.
Whereas the general expression is too complex to directly extract
relevant information we restrict ourselves to the limits $Z \gg
\Theta$ and $Z \ll \Theta$. Since we always consider the case
$\Theta \gg 1$ we  express the final result as a linear expansion
in $1/\Theta$.

\subsection{The limit of large disorder $Z \gg \Theta$ for $\alpha \Theta \ll 1$}

We start by solving this set of equations, i.e. consider a single
long-energy site. In the present case we have the two small
parameters $1/\Theta$ and $\Theta/Z$. In practice we take the
exact solution of this set of linear equations, expand
this expression in powers of $1/Z$ and then, for fixed order $1/Z$, expand in $1/\Theta$.
Finally, one keeps those up to linear order in $1/\Theta$ and $\Theta/Z$.

After a straightforward but somewhat lengthy calculation one obtains
\begin{eqnarray}
q_{central} & =& c(y,Z) Z^2  q_{rest},\label{central} \\
q_{ortho} &=&  \frac{1}{Z^2}q_{central}, \label{ortho} \\
q_{left} &=&  \frac{y^{-2}}{Z^2} \left [ 1 +
\frac{\Theta}{Z}(y-y^{-1}) \right ] q_{central} , \label{left} \\
q_{right}& = & \frac{y^{2}}{Z^2} \left [ 1 -
\frac{\Theta}{Z}(y-y^{-1}) \right ] q_{central} , \label{right}
\end{eqnarray}
with
\begin{equation}
c(y,Z) = 1 - \frac{1}{\Theta}\frac{(y - y^{-1})^2}{f(y,Z)} .
\label{cy}
\end{equation}
where
\begin {equation}
f(y,Z) =  1 + \frac{\Theta}{Z}(y + y^{-1}-1).
\end{equation}
The factor $f(y,Z)<1$ is close to unity and keeps track of
corrections due to finite disorder as expressed by $Z$. Note that
for finite field, i.e. $y \ne 1$, one obtains $c(y) < 1$. This
relation is of key importance for the subsequent discussion.
Naturally, for vanishing external field, i.e. $y = 1$, the
Boltzmann populations are recovered.

Eqs.\ref{ortho}-\ref{right} have a simple physical interpretation
in the limit of large disorder, e.g. when the correction due to
the $\Theta/Z$-term can be neglected. If a particle is sitting
adjacent to a low-energy site the probability to jump to the
low-energy site is by a factor of $Z/\Theta$ larger than to jump
to one of the other $\Theta+1$ neighbor sites. Thus, many
forward-backward jumps will occur between the low-energy site and
its neighbors. As a consequence {\it local thermodynamic
equilibrium} is generated around the central site which gives
rise to the notation of a trap-region. The $y^{\pm 2}$-factors
express that the external field attemps to shift the particle to
the right.  We mention in passing that this term has been also
used in the work of Riess and Maier \cite{Maier2} about nonlinear
conductivity, but with a very different meaning.

The factor of $c(y,Z) < 1$ captures the absolute populations in
the trap-region as compared to the population of the
neighbors. As discussed in Appendix A its value (Eq.\ref{cy}) can be also rationalized via straightforward flux arguments.

The analysis, presented so far, was obtained under the additional
assumption that all adjacent sites have the same population
$q_{rest}$. Strictly speaking this is not true because the
external field induces  an anisotropic modification of the
population. However, as discussed in Appendix
 B, this assumption
can be relieved in the regime of high dimensions and $q_{rest}$
can be simply interpreted as the {\it average} population of the
sites adjacent to the tagged region. Note that the types of sites,
which allow one to obtain this conclusion, only exist for $d \ge
3$. Thus, one might expect that major deviations are present for
$d=2$. It turns out (see Appendix B) that all higher-order terms
in the expansion of $1/\Theta$ (starting with $1/\Theta^2$) would
depend on the individual (unknown) populations. Exactly in the
limit of high dimensions they can be neglected. This justifies
that it is reasonable to stop the expansion in $1/\Theta$ after
one term. Interestingly, already this first term captures
many of the features, obtained for the numerical solution in 3D (see below).

Of particular relevance is the variation of the population of the
central site $q_{central} = q_{rest} c(y,Z) Z^2$. Due to the factor
$Z^2 \gg 1$ even a small decrease of $c(y,Z) $ from unity gives rise
to a significant decrease in population of the central site, i.e.
particles less likely populate the low-energy site. This
population difference has to be redistributed over the whole
system, i.e. to the high-energy sites.  We note in passing that
this phenomenon resembles the Poole-Frenkel effect, albeit for
different physical reasons \cite{Frenkel}. After escaping
from the trap-region as a consequence of the external field these
particles can now contribute to the total current. Since this
population increase is induced by the external field, this effect
contributes to the nonlinear response of the total system.

Now we generalize our analysis to a finite fraction $\alpha$ of
low-energy sites. One can assume that they act independently if
the probability that two low-energy sites are neighbors is small.
Formally, this corresponds to $\alpha \Theta \ll 1$.

We choose the normalization such that the average population of a
site is just unity, i.e. $ (1 - \alpha(3+\Theta)) q_{rest} +
\alpha( q_{central} + q_{left} + q_{right} + \Theta q_{ortho}) =
1$. In the limit of large $Z$ a straightforward evaluation of this
expression together with Eqs.\ref{central}-\ref{right} yields
\begin{equation}
q_{rest} = \frac{1}{1 + c(y,Z) \alpha Z^2} \label{norm1}.
\end{equation}
After an expansion in $1/\Theta$ this can be rewritten as
\begin{equation}
q_{rest} = \frac{1}{1 + \alpha Z^2} \left ( 1 + \frac{\alpha Z^2}{
\Theta (1 + \alpha Z^2)f(y,Z)} (y-1/y)^2 \right ). \label{qhigh}
\end{equation}
The current (normalized per site), as generated by the sites
beyond the trap-region is simply given by $q_{rest} (y-1/y)$.
Furthermore, one can also calculate the contribution of the
trap-regions. Although the population may be extremely high in
this region it turns out that the current of the trap-region, i.e. $q_{left}
(Zy-1/y) + q_{central}(y-1/y)/Z + q_{right} (y-Z/y) \propto \alpha q_{rest} Z^0 $ is very small
due to the factor of $\alpha$ and the missing factor of $Z$.

Inserting
Eq.\ref{qhigh} in the expression for the current and evaluation of
the first-order and third-order terms in $w$ yields after a
straightforward calculation (see Appendix C)
\begin{equation}
\label{finalmin} \sigma_{r} = 1 + \frac{\alpha Z^2}{(1 + \alpha
Z^2){f(y=1,Z)}} \frac{24}{\Theta}.
\end{equation}
Naturally, for large $\Theta$ one recovers the mean-field limit
$\sigma_r = 1$.

\subsection{The limit of small disorder $Z \ll \Theta$ for $\alpha \ll 1/\Theta$}

Using the same set of rate equations (Eq.\ref{rate1} to Eq.\ref{rate4}) as in the previous case one directly
 performs an expansion
in $1/\Theta$. One obtains after a straightforward calculation
\begin{eqnarray}
q_{central} & = &  Z^2  q_{rest} + O(1/\Theta^2), \\
q_{ortho} &= & q_{rest} + O(1/\Theta^2),
\label{ortho2}\\
q_{left} & = & [1 - \frac{1}{\Theta} (y - 1/y) (Z-1)] q_{rest},
\\
q_{right} & = & [1 + \frac{1}{\Theta} (y - 1/y) (Z-1)] q_{rest}.
\end{eqnarray}
Note that in contrast to the limit of large disorder the total population in the trap-region is not changed upon application
of an external field. In agreement with the previous case $q_{left}$ is somewhat decreased (because $Z > 1$) whereas $q_{right}$ increases.
In analogy to the limit of large disorder normalization leads to $q_{rest} = 1/(1+\alpha(Z^2-1))$.

For the current, related to one left, one central and one right site one obtains after a short calculation
$j_{left,center,right} = [(1+2Z)-(1/\Theta)(y+1/y)(Z-1)^2](y - 1/y)q_{rest}$.
The total current can be written as
\begin{equation}
j  =  \alpha j_{left,center,right} + (1-3\alpha) (y-1/y) q_{rest}.
\end{equation}
This can be rewritten as
\begin{equation}
j = \frac{1+2\alpha(Z-1)}{1+\alpha(Z^2-1)} \left [ 1 - \frac{1}{\Theta} \frac{\alpha(Z-1)^2}{1+2\alpha(Z-1)} (y+1/y)\right ](y-1/y).
\end{equation}
With a calculation in analogy to Appendix C one ends up with
\begin{equation}
 \sigma_{r }(\alpha) = 1-\frac{6}{\Theta}\frac{\alpha (Z-1)^2}{1+2\alpha(Z-1)}.
\label{sr1}
\end{equation}
In contrast to strong disorder the nonlinearity decreases as compared to the homogeneous case.
Qualitatively, this is due to the fact that upon increasing the external field intensity is shifted from the left to the right
site. This shift of population decreases the total current because a particle on the right side is more likely shifted back to the central site and thus contributes with a negative sign.

\subsection{The limit $1-\alpha \ll 1/\Theta$}
Here we deal with the limit that most sites are high-energy states. Actually, this limit is already implicitly contained in the previous calculation.
The same approximations would have been valid if $0 < Z < 1$. If, for the moment, we interpret $\alpha$
as the number of high-energy states the value of $\sigma_r$ is again given by Eq.\ref{sr1}. To keep the original notation we have to substitute
$\alpha$ by $1-\alpha$ and $Z$ by $1/Z$ in Eq.\ref{sr1}. This yields
 \begin{equation}
 \sigma_{r}(\alpha) = 1-\frac{6}{\Theta}\frac{(1-\alpha) (Z-1)^2}{Z^2+2Z(1-\alpha)(1-Z)}.
\label{sr2}
\end{equation}
Thus, independent of the degree of disorder one obtains a reduction of the nonlinearity if most states are low-energy states. The explanation is analogous to before, just the signs have to be reversed. To stress the qualitative picture one may imagine one inaccessible high-energy site. The external field pushes the particles in front of this site. Due to the inaccessibility the intensity $q_{left}$ is increased. Since, this left site cannot contribute to the current in field direction, there is an effective reduction in the total current.

\subsection{The general case except for the trap-regime}

Here we introduce a very different approach for the high-dimensional case which is applicable for the whole parameter regime
except for the trap-regime and is more general than the solutions, discussed above. The key idea is to use a statistical
description which is as close to a mean field description (infinite $\Theta$) as possible
but still reflects the properties of finite dimensions.  Generally, two approximations
reflect this limit. (1) After an orthogonal jump the previous site is randomly given a new
energy based on the energy distribution, i.e. the memory on the previous site is lost.
(2) For jumps along the field direction more information has to be
included because these jumps determine the value of $\sigma_r$. Here the memory is
lost after two successive jumps in one direction because then
the backjump probability is proportional to $1/\Theta^2$.

This statistical picture is expected to break down in the case of a trap-regime ($\alpha \ll 1$ and $Z \gg \Theta$). Here after an orthogonal jump
from a low-energy to a high-energy site it is essential to know that the previous site was a
low-energy site because very likely the system will jump back. Within the statistical
description, however, the backjump probability would be  very small because the
energy of the previous site would be determined according to the energy distribution
which would, very likely, change the energy and thus suppress the trapping property.

The formal description, resulting from (1) and (2) requires the
introduction of a probability $p_{ijk}$ that a particle sits on
site $j \in\{+,-\}$ and has neighbor sites with energies $i$ and
$k$ anti-parallel and parallel to the field direction,
respectively ($i,k \in\{+,-\}$). Then the corresponding master
equation reads

\begin{eqnarray}
  \dot{p}_{ijk} &=& \sum_h \alpha_k\, p_{hij} y \gamma_{ij} + \sum_l \alpha_i\, p_{jkl}
    \frac{1}{y} \gamma_{kj} + \Theta\,\alpha_i \alpha_j \alpha_k \left(\sum_{\tilde{i},\tilde{k}} p_{\tilde{i}-\tilde{k}} \gamma_{-j} + \sum_{\tilde{i},\tilde{k}} p_{\tilde{i}+\tilde{k}} \gamma_{+j} \right) \nonumber \\
  &&\nonumber \\
  &-& p_{ijk} \left(y \gamma_{jk} + \frac{1}{y} \gamma_{ji} + \Theta  \left( \alpha_-\gamma_{j-} + \alpha_+\gamma_{j+} \right)\right ) .
\label{formel:mod_mastergl}
\end{eqnarray}

For the motivation of the individual terms we start by discussing
the first term. The rate of jumping from site $i$ to site $j$ in
the direction of the field is proportional to $p_{-ij}+ p_{+ij}$,
which is just the probability that two adjacent sites $(i,j)$ do
exist. Since we want to calculate the change in the probability
$p_{ijk}$ we also have to take into account that statistical
factor $\alpha_k$ to have an adjacent site $k$. Furthermore a
factor of $y$ keeps track of the direction of the jump. In
summary, we obtain a contribution of $\sum_h \alpha_k\, p_{hij} y
\gamma_{ij}$. To calculate the gain term after an orthogonal
transition (third term) one has to keep into mind that the rate
only depends on the energy difference of the old site (which was
either plus or minus) and the energy of the new site $j$.
Furthermore, the product $\alpha_i \alpha_j \alpha_k$ just
expresses the probability to find three adjacent sites $(i,j,k)$.
As discussed under point (1), no memory about  the old site (plus or minus)
is kept. The remaining terms can be derived analogously.

For the calculation of the stationary limit this corresponds to a
set of 8 coupled linear equations. Again we perform a $1/\Theta$
expansion of the resulting population properties and, correspondingly,
the resulting current. Here we just present the
final result which we have checked by solving these equations by
MAPLE. In lowest order of $1/\Theta$ the current reads
\begin{equation}
\label{moda}
 J(\alpha,w,Z,\Theta) =\frac{(1-\alpha+\alpha Z)^2}{1-\alpha +\alpha Z^2}\left[ 1 - \frac{1}{\Theta}\,(y+1/y)\,\frac{\alpha(1-\alpha)(Z-1)^2}{(1-\alpha + \alpha
Z)^2}\right] (y-1/y).
 \end{equation}

 This directly yields
\begin{equation}
 \sigma_{r}(\alpha) \approx 1-\frac{1}{\Theta}\,6\,\frac{\alpha(1-\alpha)(Z-1)^2}{(1-\alpha + \alpha Z)^2}.
\label{formel:sigma_r_alpha}
\end{equation}
Most importantly, this general relation has Eq.\ref{sr1} and Eq.\ref{sr2} as limiting cases
if choosing $\alpha \ll 1 $ or $1-\alpha \ll 1 $, respectively. As described above,  the trapping picture cannot be described by this ansatz.

\section{Comparison of the numerical and the analytical results}

The key results of this work deal with the limits $\alpha \Theta \ll 1$ or $\alpha \Theta \gg 1$ for the physical relevant limit of high disorder, i.e.
$Z \gg \Theta$. In a first step we compare the numerical results for $d=3$ and $d=6$ for $\sigma_1$ and $\sigma_r$  with the analytical expressions in the respective limits. As before, we use $Z = \exp(4)$ and choose $\sigma_r - 1$ for our comparison.

\begin{figure}[tb]
\centering
\includegraphics[width=0.45\columnwidth]{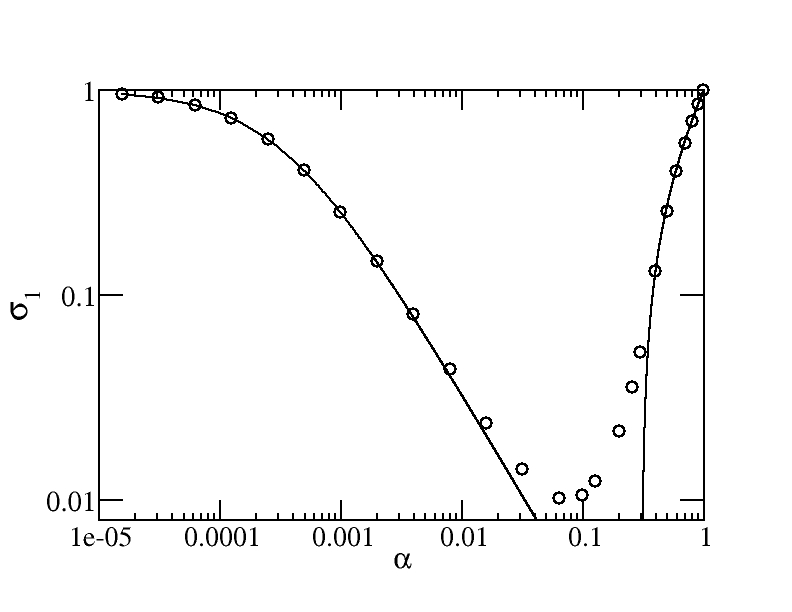}
\includegraphics[width=0.45\columnwidth]{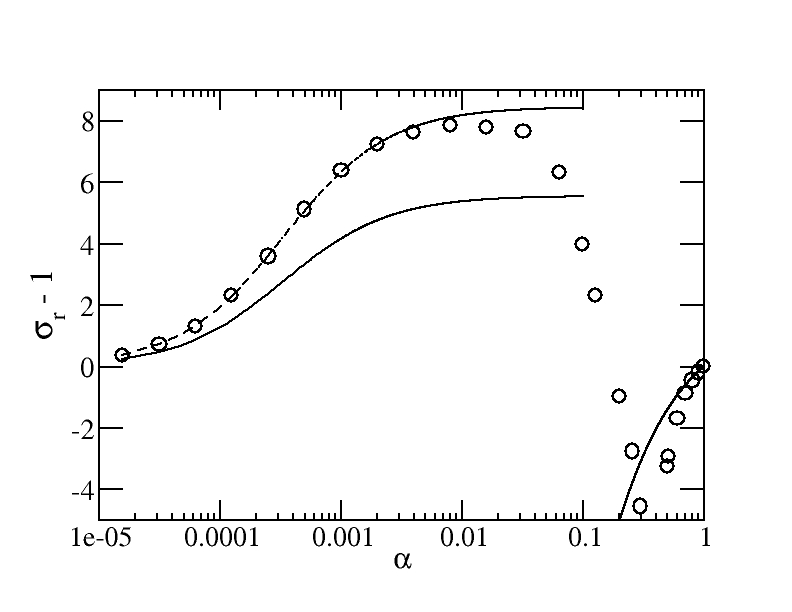}
\caption{Squares: Numerical results for $\sigma_1(\alpha)$ (left) and $\sigma_3(\alpha)$ (right) for the 3D case. Solid lines: Analytical solutions for both the limits of small and large $\alpha$. The value of $Z$ is chosen as $\exp(4)$. Broken line: The analytical solution is scaled by a factor of 1.5.} \label{figc1}
\end{figure}

The results for the 3D case are shown in Fig.\ref{figc1}. The linear conductivity $\sigma_1$ is reproduced very well. Qualitatively, $\sigma_r - 1$ is also reproduced for both limits of small $\alpha$ and large $\alpha$. Interestingly, when scaling the theoretical curve by a constant factor of 1.5 one achieves a perfect agreement between numerics and analytical theory  for $\alpha \le  2 \cdot 10^{-3}$. This scaling property is just a consequence of the fact that in this regime the individual traps act independently. However, the fact that also the sigmoidal shape of $\sigma_r-1$ is reproduced clearly shows that the trap-mechanism, as discussed above, fully explains the nature of the nonlinear conductivity in the experimentally relevant 3D case. Also the $\alpha$-dependence for large $\alpha$, giving rise to negative $\sigma_r - 1$, is reproduced. However, a quantitative comparison is only  possible for $\alpha \ge 0.9$.

\begin{figure}[tb]
\centering
\includegraphics[width=0.45\columnwidth]{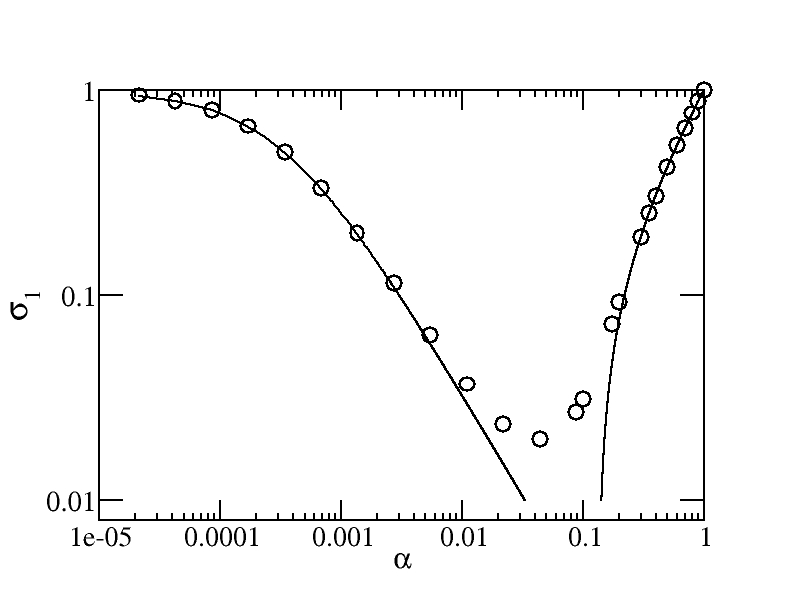}
\includegraphics[width=0.45\columnwidth]{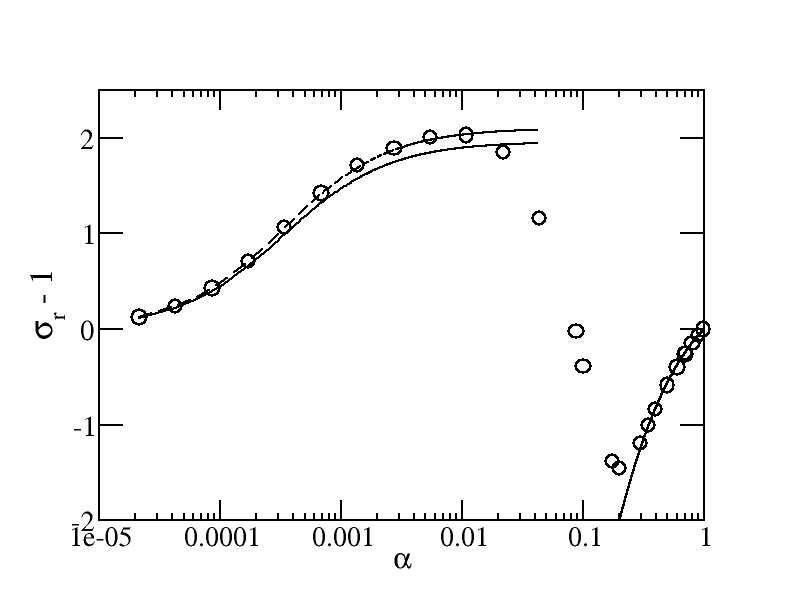}
\caption{Same as the previous figure for the 6D case. The analytical solution is scaled by a factor of 1.08. } \label{figc2}
\end{figure}

Next we compare the numerical and analytical results for $d=6$; see Fig.\ref{figc2}. In agreement with the 3D case the linear conductivity is reproduced very well in both limits. The crossover regime has a similar size (one order of magnitude on the $\alpha$-scale) but is shifted to somewhat lower $\alpha$-values. This is to be expected because the crossover should scale like $1/\Theta$. Most remarkable, the agreement for the non-linearity $\sigma_r - 1$ is very good. The trap-regime is fully reproduced, and the remaining scaling factor is reduced from 1.5 in 3D to 1.08 in 6D. This improvement does not come as a surprise because the analytical theory should be particulary good for high dimensions. A dramatic improvement as compared to the 3D case is observed for the barrier-regime. Here a quantitative agreement is seen for $\alpha$ as small as 0.3. Obviously, the neglect of back-jump memory is by far more justified for the 6D than for the 3D case.

\begin{figure}[tb]
\centering\includegraphics[width=0.45\columnwidth]{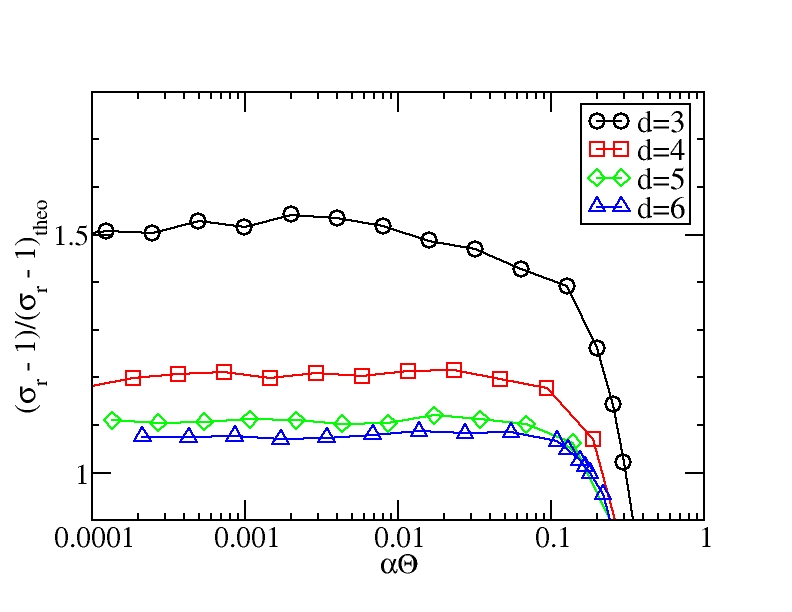}
\includegraphics[width=0.45\columnwidth]{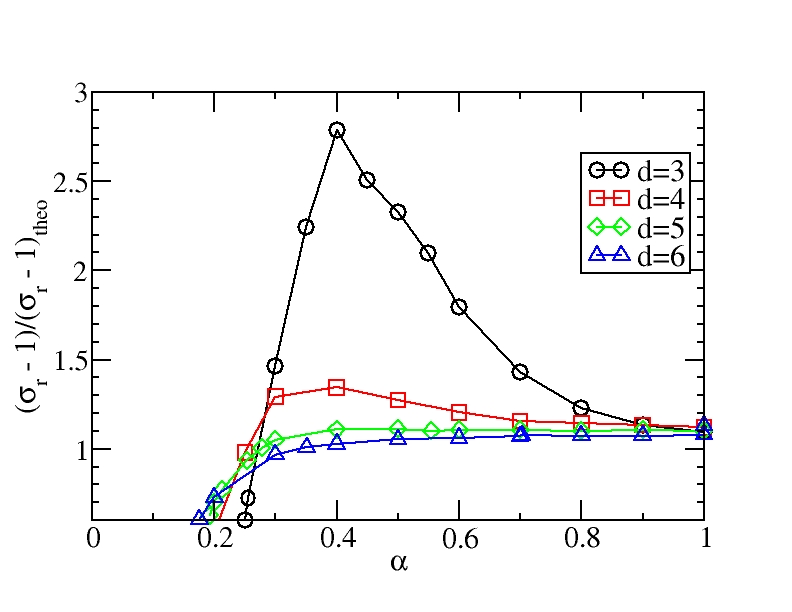}
\caption{The ratio of the actual numerical and the approximate theoretical value of $(\sigma_r-1)$ for small $\alpha$ (left) and large $\alpha$ (right). The lines serve as a guide for the eyes. } \label{figcall}
\end{figure}

For both dimensions we observe that the breakdown of the theoretical expectation for $\sigma_1$ and $\sigma_r$ in the trap-regime occurs at very similar values of $\alpha$ (3D: $\alpha \approx 0.04$; 6D: $\alpha \approx 0.02$). We checked that the same holds if we directly compare $\sigma_3$ with the theoretical expectation. Thus, the range of applicability of the linear and nonlinear prediction is very similar.

Finally, we compare the ratio of the actual numerical and the approximate theoretical value of the nonlinearity parameter $(\sigma_r-1)$; see Fig.\ref{figcall}. This representation highlights the results, discussed in Figs.\ref{figc1} and \ref{figc2} and, furthermore, allows one to see the gradual increase of predictability when going beyond the 3D case. In the trap-regime (small $\alpha$) we compare the different dimensions as a function of the scaled variable $\alpha \Theta$. It turns out that for all dimensions a plateau-value emerges. As discussed above this implies the relevance of the trap-picture for all dimensions. The plateau-value strongly depends on the dimension and approaches one for high dimensions. As discussed before, these deviations reflect that quality of the approximation that $q_{rest}$ can be identified with the average population. It becomes good for high dimensions. Interestingly, in all cases the plateau-behavior breaks down for $\alpha \Theta \approx 0.1$. As mentioned above, the presence of non-adjacent low-energy states and thus the applicability of the analytical approach in the trap-regime was related to the inequality $\alpha \Theta \ll 1$. In the opposite case, i.e. in the barrier-regime, the analytical prediction is good for $\alpha \ge 0.3$ except for the 3D case where the prediction already fails for $\alpha < 0.8$. As expected from the range of validity of the underlying approach, the breakdown of this approximation is related to $\alpha$ rather than $\alpha \Theta$, at least for $d \ge 4$. 

We also performed simulations in the limit $1 \le Z \le \Theta$ and $\alpha \Theta   \ll 1$, as reflected by Eq.\ref{sr1}. Although the qualitative features, in particular a negative $\sigma_r - 1$, can be reproduced, a quantitative comparison requires very high dimensions which are beyond the scope of our computer simulations.

\section{Discussion}

\begin{figure}[tb]
\centering\includegraphics[width=0.8\columnwidth]{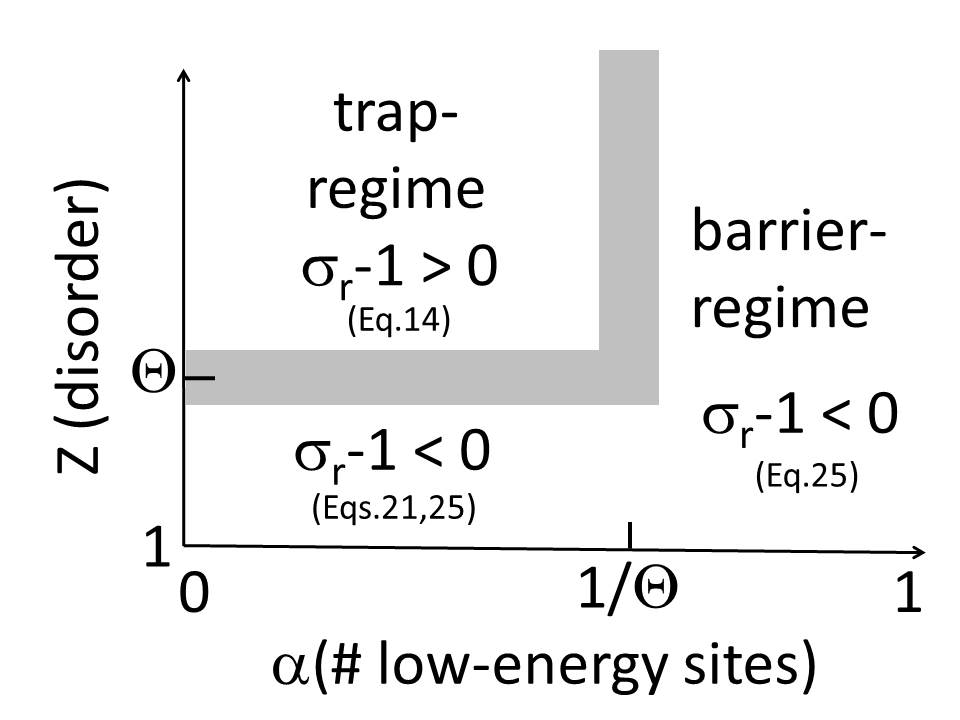}
\caption{The nonlinear response in dependence of $\alpha, Z$ and
$\Theta$ as discussed in this work. } \label{fig_schema}
\end{figure}

In this work we have described the nonlinearity for a simple
lattice model with a bimodal energy distribution in dependence of
the fraction of low-energy sites $\alpha$, the disorder (via $Z$)
and the dimension (via $\Theta$).  A clear physical picture is presented for the
limits of a small number ($\alpha \ll 1/\Theta$) and a large number ($1-\alpha \ll 1$) of low-energy states in the limit of large disorder ($Z \gg \Theta)$; see Fig.\ref{fig_schema} for a sketch of the whole $\alpha,Z$-phase space.
In the first case the few low-energy states serve as traps which are in a local thermodynamic equilibrium-state with the adjacent
sites. This gives rise to a reshuffling of intensity to the high-energy states so that the positive nonlinear contribution is via the nonlinear increase of the current of the high-energy states, i.e. $\sigma_r > 1$. In the other limit of many low-energy states the remaining high-energy states serve as barriers which accumulate intensity on the neighbor site opposite to the field direction and decreases the intensity on the adjacent site parallel to the field direction, respectively. The contribution to the total current, however, has the reverse order. Thus, one obtains an effective decrease of the non-linear conductivity, i.e. $\sigma_r < 1$.

Note that for sufficiently large $Z$ there exists a regime where $\sigma_r = 1 + 24/\Theta$ does not depend on $\alpha$ ($1 \ll \alpha Z^2$). Whereas this theoretical plateau value is close to the numerical value for higher dimensions, a difference of a factor of 1.5 is observed for the 3D case (see Fig.\ref{figc1}). Furthermore, in this limit also the $Z$-dependence is very weak (only via $f(y,Z)$) and disappears for large $Z$.
We mention in passing that also the 2D model displays a plateau value. However, here the numerical value is even two times higher than the theoretical value (data not shown).

For large disorder the theoretical predictions are not applicable in the regime $0.1 < \alpha  \Theta < 2.5$ (see Fig.\ref{figcall}).  Note that the critical value $\alpha_c$ for site percolation for high dimensions is approximately $1/\Theta$ (e.g.
$\alpha_c(d=13) \approx 1.025/\Theta$ \cite{Grassberger}). It is
known that above the percolation threshold the linear
conductivity strongly increases because the particles can cross
the sample without using the high-energy states. The present
results show that also the nonlinear conductivity shows a
significant change around this critical concentration of
low-energy sites. However, a closer relation of the crossover-regime to the
percolation approach is beyond the scope of this work.

From a theoretical perspective one might want to consider more
complex energy distributions. Although the general physical
mechanisms, described in this work, may remain the same, a
quantitative analysis would become much more difficult if not
intractable within the present approach. E.g., the analysis of
a model with three instead of two different energy levels in the barrier-regime would
require the solution of 27 instead of 8 linear equations.
Nevertheless,  the insight of the present work on the two-state
model can be used to rationalize the numerical results, previously
obtained for hopping models with continuous energy distributions
(see Introduction). For a Gaussian distribution the resulting
nonlinearity was analyzed in dependence on the lowest energy
$U_{min}$ of all sites \cite{ZPC}. Among the different
realizations of the Gaussian random numbers the value of $U_{min}$
fluctuates quite significantly. It turned out that (1) only for
sufficiently low values of $U_{min}$ one indeed observes $\sigma_r
 > 1$ whereas otherwise one has $\sigma_r < 1$ and (2) in
the regime of sufficiently low values of $U_{min}$ (i.e.,
$\sigma_r  > 1$) the value of $\sigma_r$ hardly depends on the
specific value of $U_{min}$ \cite{ZPC}. Both observations can be
qualitatively explained by the present results. Due to the
dependence on the single low-energy value $U_{min}$ the continuous
distribution can be related to a discrete model with one
particular low-energy state. Qualitatively, this corresponds to the
present bimodal model in the trap-regime. If $\Delta U$ is too small one
indeed observes $\sigma_r < 1$ because local equilibration does
not occur and the nonlinearity is explained via the barrier-mechanism. Furthermore, for large
$Z$ no dependence on the disorder is observed (explanation of
observation(2)). Thus, several non-trivial observations from a
continuous Gaussian distribution can be rationalized by the
insights from the bimodal model. In previous simulations also the
box-type distribution was analysed somewhat closer. It was shown
that in this case one observes $\sigma_r < 1$ \cite{ZPC}. When
mapping this distribution on a bimodal distribution an appropriate
choice is $\alpha \approx 1/2$. Here we obtain $\sigma_r < 1$ in agreement with the numerical result.

The rate equations Eq.\ref{rate1} to Eq.\ref{rate4} required implementation of the
approximation that the population of all neighbor sites is identical. Although we found
a strict argument why this approximation is not required for the validity of our approach, it is promising
that the general approach Eq.\ref{formel:mod_mastergl} yields identical results for the limits $1-\alpha \ll 1/\Theta$ as well as $\alpha \ll 1/\Theta$ together with $Z \ll \Theta$. This serves as an important consistency check of the two complementary approaches, used in this work.

As outlined in the Introduction hopping models constitute a
reasonable approach for the real-world situation in inorganic ion
conductors. The previous numerical results indicated that the experimental observation $\sigma_r > 1$ suggests a very broad distribution of the site energies such that there exist a few sites with very low energies \cite{ZPC}. However,
in the previous work no physical picture for this observation
could be supplied. The present work suggests that the presence of
these low-energy sites gives rise to the phenomenon of local
thermodynamic equilibrium which is responsible for the
experimental observation $\sigma_r > 1$. Please note that the
terminology of sites was related to the vacancy picture. Going
back to the ion rather than the vacancy picture the positive
nonlinear effect translates into the presence of a few unfavorable
sites which nevertheless have to be frequently visited by the ion
due to the shortage of vacant sites.

In summary, to the best of our knowledge we have for the  first time
formulated quantitative predictions for the non-linear response of
a disordered hopping model in high dimensions with several
predictions which semi-quantitatively already hold for the
experimentally relevant 3D case and hold very well for even higher dimensions. Thus, the experimentally
observed positive nonlinear effect indicates the presence of
strong disorder with only a few sites with very low energy as
outlined for the case of the Gaussian distribution. A more direct
experimental analysis of non-linear single-particle dynamics is
possible with modern holographic-mechanical optical tweezers
\cite{Egelhaaf,Egelhaaf2} which can generate random potentials and
may allow for a more quantitative comparison with the dynamics in
disordered energy landscapes.

We acknowledge very helpful discussions with  B. Roling,  S.
R\"othel, T. Franosch, and S. Egelhaaf about this project. In particular we
would like to acknowledge the very fruitful collaboration with our
late colleague R. Friedrich on the topic of this work. We
gratefully acknowledge the financial support by the DFG via FOR
1348 and SFB 458.

{\bf Appendix A}

Here we present an alternative argument for Eq.\ref{cy} in the limit $\Theta/Z \rightarrow 0$.
 In the stationary
non-equilibrium case the total fluxes in and out of the
trap-region have to be identical. Counting the number of possible
transitions the flux balance can be written as
\begin{equation}
\label{eqflux}
q_{left} (\Theta + 1/y) + q_{right} (\Theta + y ) + q_{ortho} \Theta (\Theta + y + 1/y - 1)
= q_{rest}(\Theta^2 + \Theta(y + 1/y + 1) + y + 1/y)
\end{equation}

Neglecting the terms of order $\Theta^0$ and using
Eqs.\ref{ortho}-\ref{right} (without the correction terms of order
$\Theta/Z$) this can be rewritten as
\begin{equation}
(y^{-2} + y^{2} - 2 + y + 1/y +1 + \Theta) q_{ortho} = (y + 1/y +1 +\Theta) q_{rest}
\end{equation}
which after expansion in $1/\Theta$ and comparison with
Eq.\ref{ortho} directly yields Eq.\ref{cy} (again without the
correction terms). Thus, the factor $c(y,Z) < 1$ balances the increase
of intensity of $q_{left} + q_{right}$, following from $y^2 +
y^{-2} > 2 $ for $y > 1$.

{\bf Appendix B}

Here we show that for the derivation of the leading term of $c(y)$ it does not matter
whether the populations of the sites, adjacent to the trap-region, are identical or not.
Using a vector notation we choose the central site as $(0,...,0)$, the left and right sites
as $(\pm 1, 0, ..,0)$ and the orthogonal sites as $(0,...,\pm1, ...)$. Coordinates different
from zero are explicitly listed .Then
the neighbors of the trap-region can be classified into six types:
(1) $(0,...,\pm1, ...,\pm1, ...)$, (2) $(0,...,\pm 2, ...)$, (3) $(+1, ..., \pm 1, ...)$, (4) $(-1, ..., \pm 1, ...)$,
(5) $(+2,...)$, (6) $(-2,...)$. Within one type all populations are identical due to symmetry reasons. The individual
populations are denoted $q_i$ with $i \in {1,...,6}$. The number of sites $n_i$ can be easily calculated. It turns out that only $n_1$ is of order $\Theta^2$ (namely $n_1 = (1/2) \Theta^2 - \Theta$) whereas all other $n_i$ are at most of order $\Theta$ $(n_2 = n_3 = n_4 = \Theta, n_5 = n_6 = 1)$ . The total number of sites $n_{tot}$ is $(1/2) \Theta^2 + 2 \Theta + 2$. Denoting $q_{rest}$ as the average population of a site one has the simple relation
\begin{equation}
\label{app1}
\sum_{i=1}^6 n_i q_i = n_{tot} q_{rest}.
\end{equation}

In the mean-field limit $\Theta \rightarrow \infty$ one can write $q_i = q_{rest}$ because the Boltzmann probabilities are recovered. For finite but large dimension one can choose the expansion ansatz $q_i = q_{rest}(1 + a_i/\Theta + b_i / \Theta^2 + ...)$ where, again, $q_{rest}$ denotes the average population of all sites. Inserting this ansatz into Eq.\ref{app1} and comparing all terms of order $\Theta$ one immediately obtains $a_1 = 0$. Thus, in case of symmetry breaking  the impact of finite order on $q_1$ is only of order $1/\Theta^2$.

Going back to the rate equations Eq.\ref{rate1} to Eq.\ref{rate4} one has to modify the source terms by replacing $q_{rest}$ by the specific populations $q_i$, defined above. After a straightforward calculation one realizes that for the relevant orders in $\Theta$ only the term $a_1$ would contribute which, however, is zero. Thus, all variations of $q_i$ from $q_{rest}$ are irrelevant for the first order expansion in $1/\Theta$ and would only show up in the higher-order terms.

Of course, the above arguments can be also applied for the next shell of sites around the sites considered above. Again, there is one type of sites which dominates all other sites in the limit of high dimensions. Reiterating this argument one can conclude that $q_{rest}$ can be even identified with the average population of all sites of the system beyond the tagged region if one restricts oneself to the corrections of order $1/\Theta$.

We mention in passing that for the 2D case ($\Theta=2$) one has $n_1 = 0$. Thus, the key states for argumentation only start to exist for $d=3$ and a quantitative comparison for the 2D case should be quite poor. This is indeed reflected by the actual comparison as indicated in the main text.

{\bf Appendix C}

For the calculation of the nonlinear conductivity in Sect. IV B we
start with the relation

\begin{equation}
q_{rest} \propto \left ( 1 + k (y-1/y)^2 \right ). \label{qhigh2}
\end{equation}
with
\begin{equation}
k = \frac{\alpha Z^2}{ \Theta (1 + \alpha Z^2)}.
\end{equation}

The current $j$ is given by $(y+1/y)q_{rest} \propto (w + w^3/6)
q_{rest}$. Up to third order in $w$ one can write (using $y =
\exp(w)$)

\begin{equation}
j \propto (w + w^3/6)( 1 + k (2w)^2) \propto w(1+(4k+1/6)w^2).
\end{equation}

This immediately yields Eq.\ref{finalmin}.


\end{document}